\documentclass[global,twocolumn,final]{svjour}
\usepackage{graphics}
\usepackage{cite}
\usepackage{verbatim}

\journalname{}

\newcommand{\ltish}{\protect\raisebox{-0.4ex}{$\,\stackrel{<}{\scriptstyle\sim}\,$}}

\begin{document}

\title{Heating rate and electrode charging measurements in a scalable, microfabricated, surface-electrode ion trap}
\author{D.T.C. Allcock\inst{1} \and T.P. Harty\inst{1} \and H.A. Janacek\inst{1} \and N.M. Linke\inst{1} \and C.J.Ballance\inst{1}  \and A.M. Steane\inst{1} \and D.M. %%@
Lucas\inst{1} \and R.L. Jarecki Jr. \inst{2} \and S.D. Habermehl\inst{2} \and M.G. Blain\inst{2} \and D. Stick\inst{2} \and D.L. Moehring\inst{2}}

\institute{Department of Physics, University of Oxford, Clarendon Laboratory, Parks Road, Oxford, OX1 3PU, UK \and Sandia National Laboratories, Albuquerque, New Mexico 87185, USA}

\date{Received: date / Revised version: date} % The correct dates will be entered by the editor

\maketitle

\begin{abstract}
\begin{sloppypar}
We characterise the performance of a surface-electrode ion ``chip" trap fabricated using established semiconductor integrated circuit and micro-electro-mechanical-system (MEMS) microfabrication processes which are in principle scalable to much larger ion trap arrays, as proposed for implementing ion trap quantum information processing.  We measure rf ion micromotion parallel and perpendicular to the plane of the trap electrodes, and find %%@
that on-package capacitors reduce this to \ltish{10\,nm} in amplitude. We also measure ion trapping lifetime, charging effects due to laser light incident on the trap electrodes, and the heating %%@
rate for a single trapped ion. The performance of this trap is found to be comparable with others of the same size scale.
\end{sloppypar}
\end{abstract}

\section{Introduction}
\label{intro}

Many of the requirements for quantum information processing have been demonstrated using small numbers of trapped, laser-cooled ion-qubits (see \cite{Wineland11} for a recent review). Two %%@
significant present challenges are to scale up these systems to large numbers of qubits, and to reduce the so-called ``anomalous heating'' affecting the ions' external motion which is used %%@
to implement quantum logic gates between neighbouring qubits \cite{Turchette00}. In this paper we report on the performance of a trap which is constructed using intrinsically scalable %%@
semiconductor fabrication technology.  In particular we measure the motional heating rate of a single ion. The fabrication process is described in more detail in \cite{Stick10}, where initial %%@
characterization of the trap is also reported. This trap is designed to be compatible with integration of microfabricated optical elements, and it has already been used to demonstrate %%@
collection of ion fluorescence using diffractive micro-optics \cite{Brady10}.  Ion shuttling through junctions has also been demonstrated in traps  utilising the same fabrication process \cite{Moehring11}.

\section{Experimental apparatus}
\label{sec:2}

\begin{figure}
\resizebox{0.5\textwidth}{!}{\includegraphics {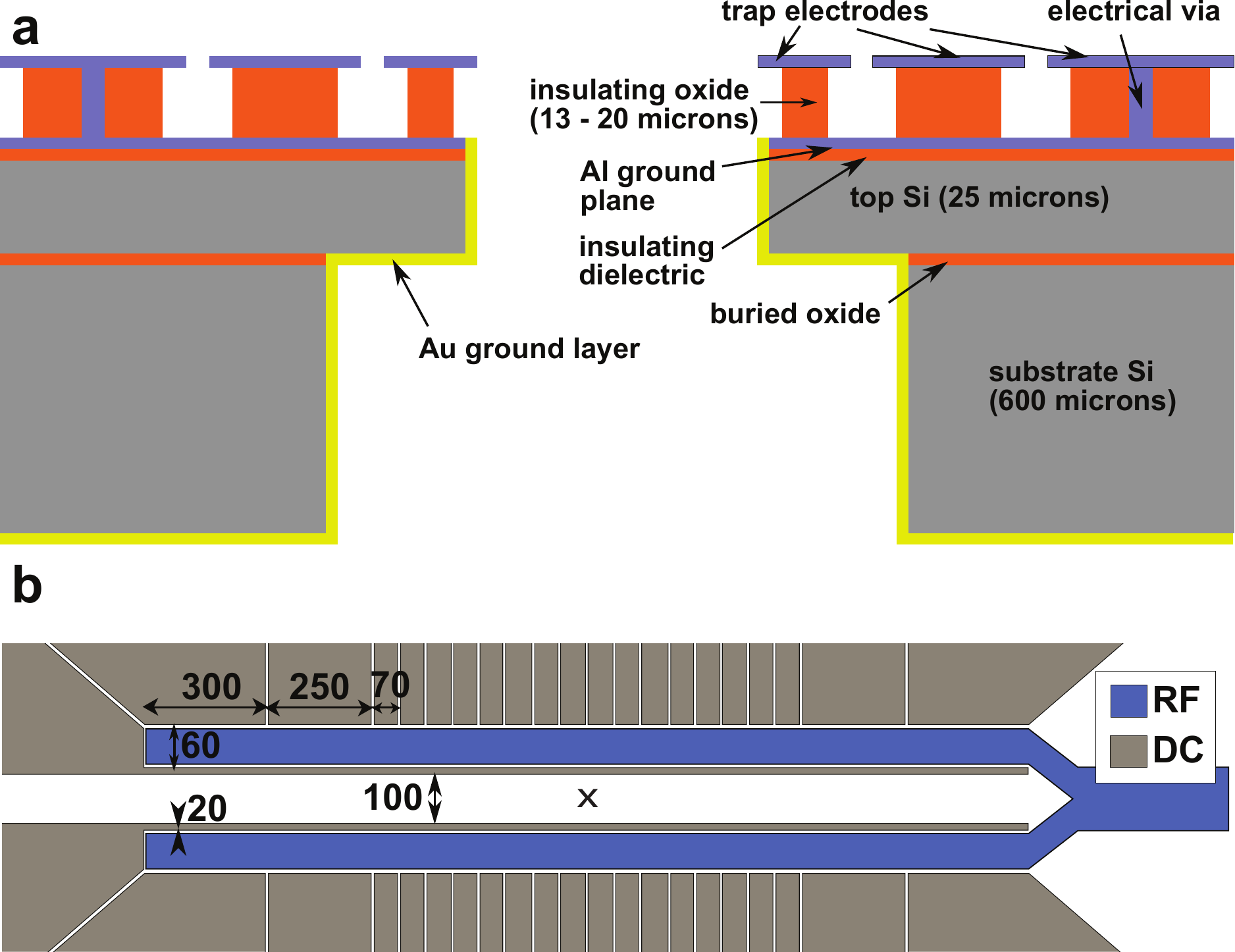}}
\caption{Schematic diagram of the ion trap. (a) Cross-sectional view (not to scale), with layer thicknesses indicated in microns. The ion is trapped 84\,$\mu$m above the plane of the trap %%@
electrodes. (b) Plan view (to scale), with dimensions shown in microns.  A $\times$ marks the position of the trap centre used in these experiments. }
\label{trapdiag}       
\end{figure}

We tested three different traps of the same type. The traps are identical to those described in \cite{Stick10} except for Trap 2 which has 13\,$\mu$m high oxide %%@
pillars supporting the electrodes rather than 20\,$\mu$m high pillars (see fig. \ref{trapdiag}).  Traps 1, 2 and 3 differ in the filtering of the dc control electrodes (see section \ref{sec:3}).  The traps were mounted in the %%@
same vacuum system used in \cite{Allcock10}.  The system was modified such that the neutral calcium oven was behind, rather than to the side of, the trap so the ions are loaded through the %%@
central slot.  The oven was also loaded with isotopically enriched calcium (10\% $^{43}$Ca, 90\% $^{40}$Ca). The vacuum pressure was $<10^{-11}$\,Torr, rising to  $1\times10^{-11}$\,Torr %%@
with the oven running.

\begin{figure}
\begin{center}
\resizebox{0.45\textwidth}{!}{\includegraphics {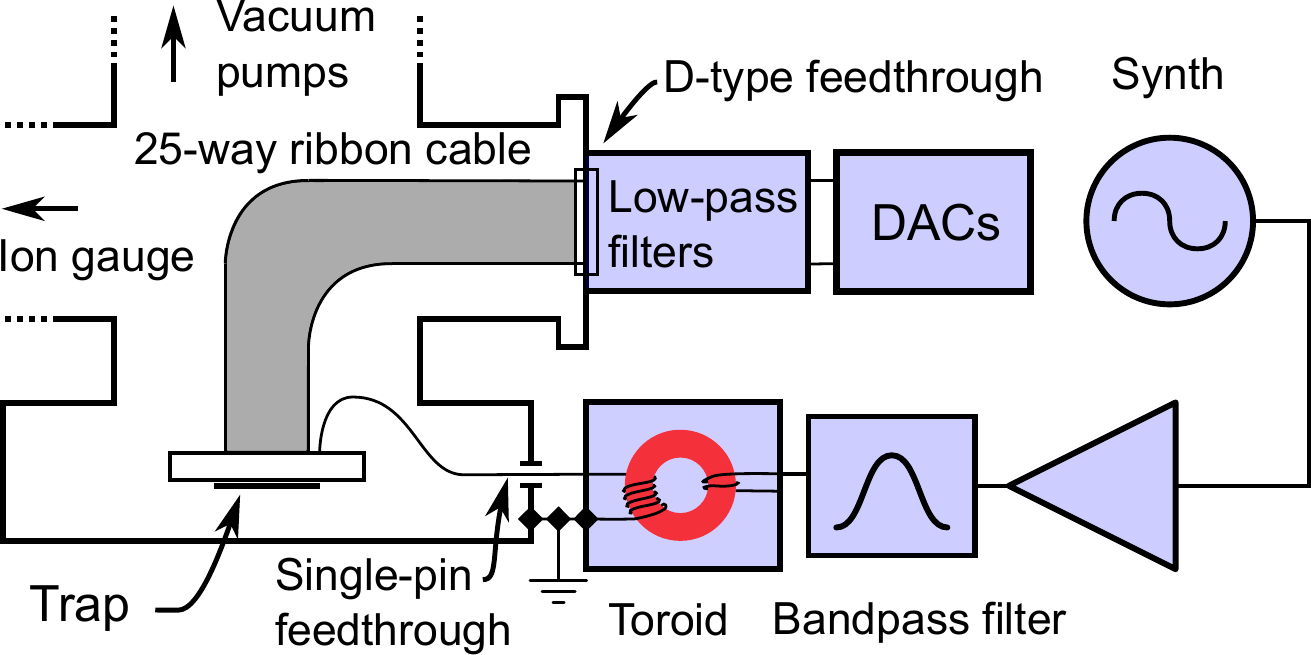}}
\end{center}
\caption{Schematic of the vacuum system and wiring arrangement of the trap. The dc electrode voltages are supplied by digital-to-analogue converters (DACs), filtered by low-pass filters, and %%@
taken to the trap by a 25-way vacuum-compatible ribbon cable (we are only using the central 10 pairs of electrodes, the rest are grounded on the package). The trap rf drive is provided by a synthesizer, which is amplified and filtered, then stepped-up using a toroid in a resonant %%@
transformer arrangement. The vacuum chamber is used as the common grounding point for rf and dc voltages.}
\label{electronicsdiag}       
\end{figure}

The rf trapping voltage is stepped-up using an iron powder toroid (Micrometals T94-6) in a resonant transformer arrangement (see fig. \ref{electronicsdiag}). The toroid gives a loaded $Q$ of %%@
28 and a resonant frequency of $\Omega_{\mbox{rf}}=2\pi\times33\,$MHz.  The actual voltage step-up at the trap is 21.8, which was calculated by measuring an ion's radial secular frequencies %%@
$\omega_r$ and deducing the trap voltage using our electric field simulation.  The voltage amplitude used was in the range 50--140\,V. 

As the toroid does not have a high enough $Q$ to filter out noise effectively at $\Omega_{\mbox{rf}}\pm\omega_{\mbox{r}}$ (which can heat the ion \cite{Wineland98}) we place a bandpass %%@
filter between the amplifier and the toroid.  The filter is a 50$\,\mathrm{\Omega}$ impedance mesh-capacitor topology with 3.2\,dB insertion loss at 33\,MHz, $-38$\,dBc at 30\,MHz and %%@
$-30$\,dBc at 36\,MHz (typical radial frequencies are 3--4\,MHz).  

DC electrode voltages of up to $\pm10$\,V are provided by an AD5372 DAC chip.  This chip has an intrinsic voltage noise of $<100$\,nV$/\sqrt{\mbox{Hz}}$ but this is reduced by a further factor %%@
of $>10^5$ by low-pass filters (RCLC topology with cut-off frequency $<$10\,Hz).  The voltages after the filter board were checked for any residual noise using a spectrum analyzer and a %%@
custom pre-amplifier with a noise floor of 0.25\,nV$/\sqrt{\mbox{Hz}}$.  As this trap has a similar split central control electrode configuration to an earlier trap tested at Oxford, we use %%@
the same technique as in that work \cite{Allcock10} to generate voltage sets with the required axial trapping frequency and radial principal axis orientation.

We use the same laser systems and collection optics as those described in \cite{Allcock10}.  Overall photon detection efficiency is 0.23\%.  This gives approximately 50,000 counts/s for a %%@
single ion cooled on the S$_{1/2}-$P$_{1/2}$ transition by one saturation intensity of resonant 397\,nm light, and repumped on both the D$_{3/2}-$P$_{3/2}$ (850\,nm) and D$_{5/2}-$P$_{3/2}$ %%@
(854\,nm) transitions. Background scatter is $\sim$100 counts/s at this 397\,nm power (2.0\,$\mu$W in a spot with $1/$e$^2$ radius $w=30\,\mu$m) \cite{Linke11}.  Using a multi-element %%@
diffraction-limited fibre output collimator (CVI-Melles Griot GLC-14.5-8.0-405) and quartz optics for the 397\,nm beam path were important in achieving this low background.  

%{$s=1$;\\ 0.5$I_{\mbox{DNS}}$}. 

\begin{figure}
\begin{center}
\resizebox{0.5\textwidth}{!}{\includegraphics {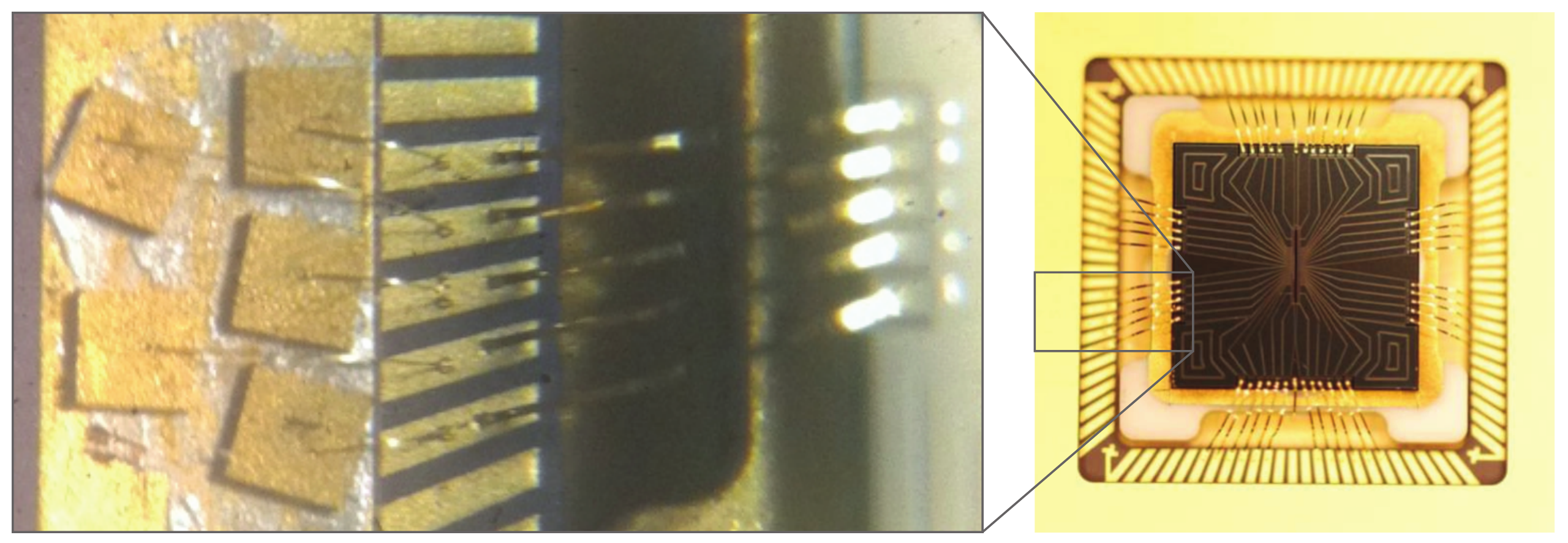}}
\end{center}
\caption{Magnified view of capacitors added to the ceramic pin grid array (CPGA). The capacitors were glued to the outer, grounded, gold ring of the CPGA, and then wire bonded to the bond pads as shown.}
\label{caps}       
\end{figure}

\begin{figure}
\resizebox{0.5\textwidth}{!}{\includegraphics {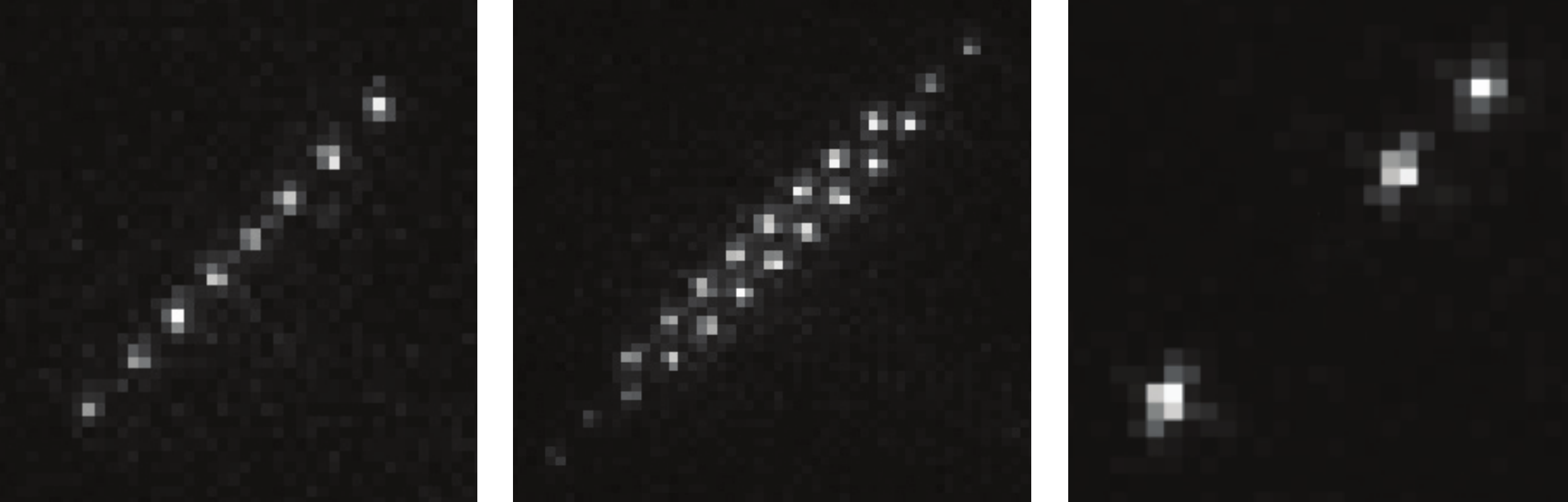}}
\caption{8 ion $^{40}$Ca$^+$ crystal (left), 21 ion $^{40}$Ca$^+$ crystal (centre) and mixed $^{40}$Ca$^+$ and $^{43}$Ca$^+$ crystal (right).  The $^{43}$Ca$^+$ ions are dark because the %%@
397\,nm cooling laser is not modulated to span the 3.2\,GHz hyperfine splitting of the transition.}
\label{ions}       
\end{figure}

\section{Micromotion compensation}
\label{sec:3}

In Trap 1 we were unable to compensate the rf micromotion with static electric fields, implying that there was no stationary rf null \cite{Berkeland98}.  The residual micromotion was mainly %%@
perpendicular to the plane of the trap electrodes, with an amplitude of $\sim$200\,nm.  This was estimated by measuring the amplitude of the micromotion sidebands on a spectrum obtained with %%@
a D$_{3/2}-$P$_{3/2}$ 850\,nm repumping beam propagating at 45$^\circ$ to the plane of the trap \cite{Allcock10}.  The direction of the motion meant that we could discount a phase difference %%@
between the two rf rails, which would cause radial micromotion parallel to the plane of the trap.  End effects due to the finite length of the trap were also predicted to be negligible from %%@
electric field simulations.  This implied that the likely cause of the micromotion was rf pickup on the dc electrodes. For this to be to be the case the pickup voltages have to be out of %%@
phase with the rf electrode voltage, and the secondary rf null created by the pickup has to be in a different place to the principal rf null.  We believe that in our case phase shifts were %%@
caused by the complex impedance of the $\sim$150\,mm long ribbon cable (see fig. \ref{electronicsdiag}) connecting the trap to the vacuum feedthrough or by crosstalk between the wires in the %%@
cable.

In an effort to reduce the micromotion in Trap 1 we added a low voltage ``compensation'' rf voltage to one of the central dc electrodes. This rf was supplied by a synthesizer phase-locked %%@
with a variable offset to the main rf drive synth.  By adjusting the phase offset and the amplitude of this synthesizer we were able to eliminate the micromotion parallel to the trap %%@
surface. This implied that a significant cause of the problem was rf pickup on these centre electrodes, which might be expected since, out of all the dc electrodes, they have by far the largest capacitative coupling to %%@
the rf electrodes and therefore will have the largest rf pickup. On Trap 2, therefore, we added a 1\,nF capacitor to the package between each of these electrodes and the trap ground plane. %%@
However, significant micromotion perpendicular to the trap surface was still present, so on trap 3 we added a 820\,pF capacitor between each dc electrode and ground.  These capacitors were %%@
glued onto the outer gold ring of the CPGA with Epo-Tek H20E conductive epoxy and then wire bonded to the pads on the CPGA (see fig. \ref{caps}). These capacitors have the effect of moving %%@
the common rf grounding point onto the trap package, eliminating differential phase-shifts due to the ribbon cable.

The addition of these capacitors reduced the residual micomotion to a similar amplitude to that caused by a 1\,V/m (10\,V/m) excess field parallel (perpendicular) to the plane of the trap %%@
(at 2.35\,MHz radial secular frequency).  This implied there was a residual motion of order 1\,nm (10\,nm), which is at the limit of what we can conveniently resolve with our compensation %%@
technique \cite{Allcock10}.  

\section{Trapping lifetime}
\label{sec:4}

Single-ion trapping lifetime was of order one hour with laser cooling and of order one minute without.  In traps 1 and 2 the two-ion lifetime was only a few minutes (with cooling) and we %%@
were unable to load large crystals.  In trap 3 the two-ion lifetime was approximately half the single-ion lifetime and we were able to load large crystals with tens of ions (see fig. %%@
\ref{ions}).  The difference in behaviour is likely to be because, with multiple ions, micromotion causes parametric heating when the ions are hot and the motion becomes anharmonic %%@
\cite{Wineland98} (for example, after a background gas collision or just after loading).  We have also successfully loaded $^{43}$Ca$^+$ ions and sympathetically cooled them using %%@
$^{40}$Ca$^+$ ions \cite{Blinov02}.

\section{Charging effects}
\label{sec:5}

The traps appear to be fairly resistant to charging under normal operation.  The field compensation in the direction parallel to the trap plane drifts by only $\sim 1$\,V/m on an hour %%@
timescale with no change during or after ion re-loading.  

\begin{figure}
\begin{center}
\resizebox{0.3\textwidth}{!}{\includegraphics {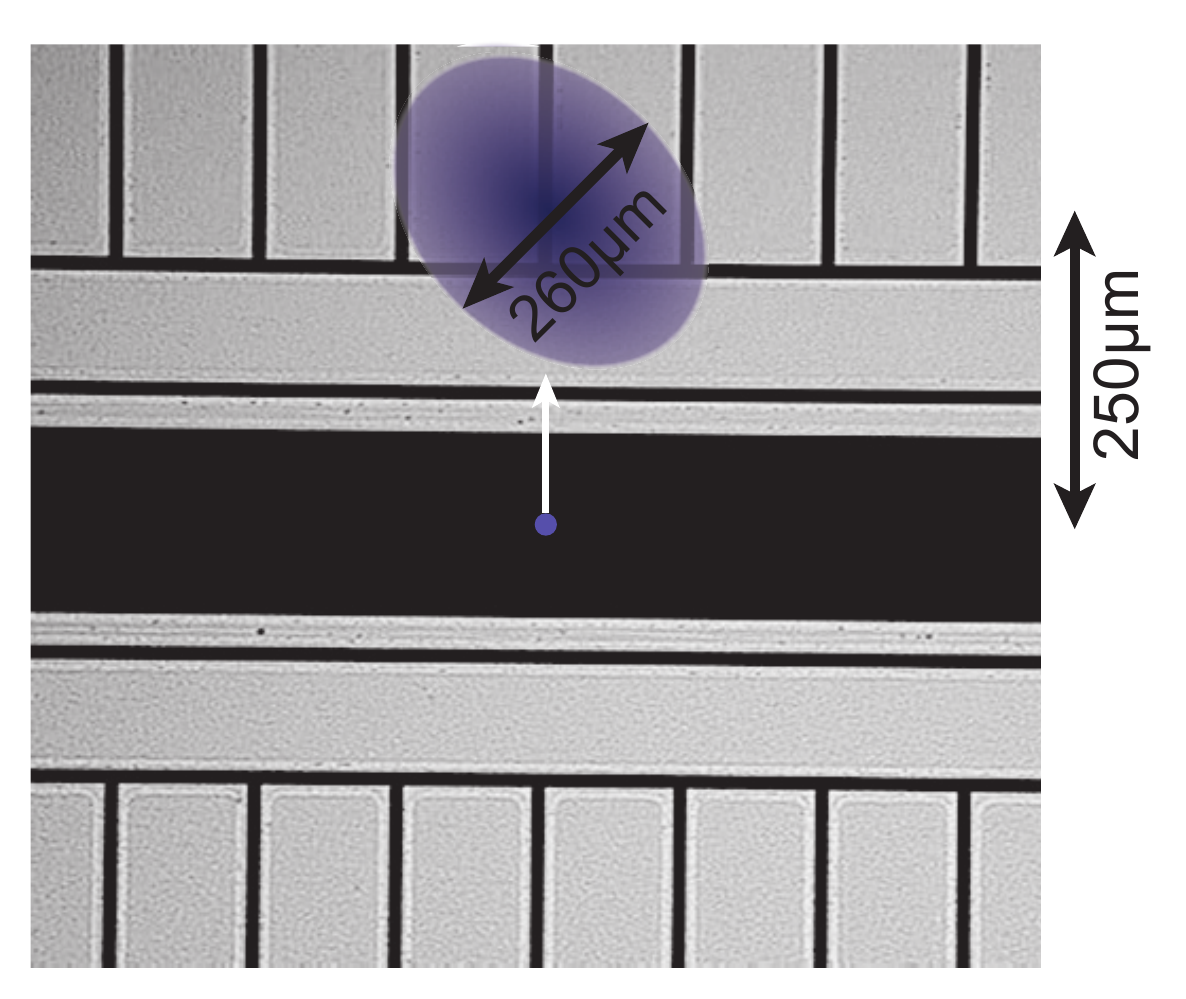}}
\end{center}
\caption{Diagram to show how the 397\,nm charging beam strikes the trap. The beam has a circular profile with spot size $w=130\,\mu$m, but propagates at 45$^\circ$ to the surface so it %%@
illuminates an elliptical area of the electrodes. The white arrow shows the direction of the induced electric field measured at the ion (blue dot).}
\label{chargingbeam}       
\end{figure}

\begin{figure}
\resizebox{0.48\textwidth}{!}{\includegraphics {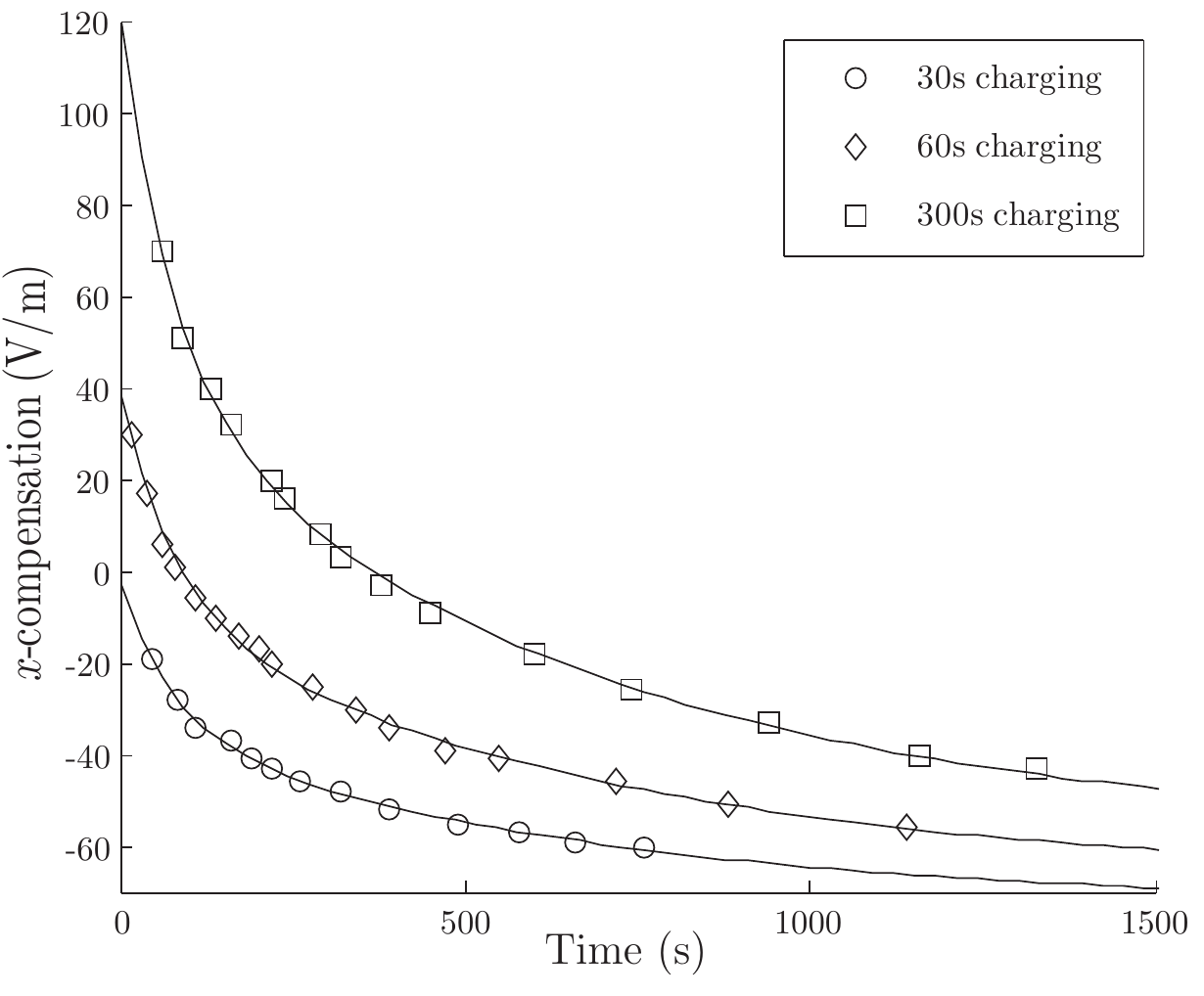}}
\caption{Electric field measured at the ion, after irradiation of the trap by 30, 60 and 300\,s exposure to the 397\,nm charging beam. The curves are a joint fit to $E_x = %%@
A_i+B_ie^{-t/t_1}+C_ie^{-t/t_2}$ and the fitted time constants are $t_1=78$\,s and $t_2=654$\,s.  Errors are approximately the size of the symbols used.}
\label{charging}       
\end{figure}

A more detailed study of the trap charging was conducted by irradiating the electrodes directly with a 397\,nm laser beam to induce a larger charging effect. The beam had a power of %%@
10\,$\mu$W in a circular spot with $1/$e$^2$ radius $w=130\mu$m and propagated at 45$^\circ$ to the surface (see fig. \ref{chargingbeam}).  Data were taken by turning the laser on for a %%@
fixed amount of time, then repeatedly recompensating the in-plane micromotion at $\sim30$\,s intervals.  This method has a resolution of $\pm$1\,V/m with 10\,s detection time.  Both the %%@
direction of the resulting field (which attracted the ion towards the beam spot) and the time dependence of its decay were in agreement with previous work by Harlander and coworkers %%@
\cite{Harlander10}. The data are well fitted by a double exponential (see fig. \ref{charging}) with time constants 77.5 and 654 seconds compared to 5 and 120 seconds reported in that work.  %%@
This is possibly because the native oxide layer on our aluminium trap was thicker or less conductive than that on the copper trap in \cite{Harlander10} and could support a longer relaxation %%@
time for any charges on it \cite{Rageh77}.  

\section{Heating rates}
\label{sec:6}

\begin{figure}
\resizebox{0.48\textwidth}{!}{\includegraphics {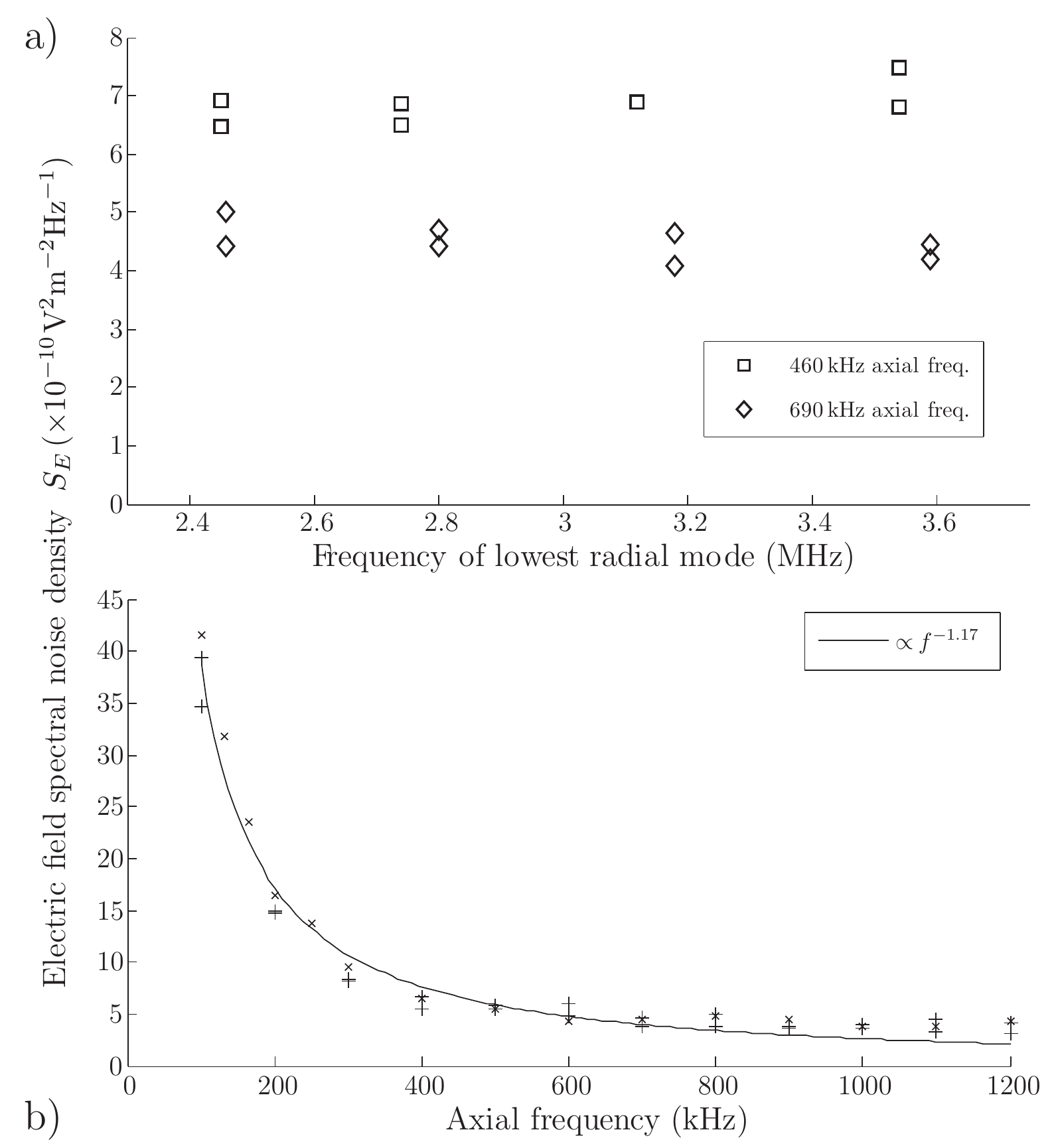}}
\caption{Single-ion heating rate in Trap 3, expressed as the electric field spectral noise density experienced by the ion. (a) Heating rate versus radial frequency at fixed axial frequencies of $f_z = 460$\,kHz and $f_z = 690$\,kHz. (b) Heating rate versus axial frequency at fixed rf amplitude ($f_{r1} \simeq 3.1$\,MHz, $f_{r2} \simeq3.5$\,MHz). The data points marked $(\times)$ were taken a week after those marked $(+)$. The curve %%@
is the least-squares fit to a $1/f^n$ power law. The heating rate equates to $\sim55$ quanta/ms at 1\,MHz axial frequency.}
\label{axialheating}       
\end{figure}

We made several studies of the heating rate in Trap 3 using our simplified method for ${\mathrm \Lambda}$ systems \cite{Allcock10} based on that described in \cite{Wesenberg07}.  We see no significant effect on the heating rate due to variations in the radial frequency (fig. \ref{axialheating}a) which implies %%@
that we are observing axial mode heating only.  The heating rate dependence on the axial frequency (see fig. \ref{axialheating}b) below $f\approx 750$\,kHz is comparable with the $1/f$ %%@
dependence reported in the literature \cite{Deslauriers06,Labaziewicz08,Turchette00}.  Above 750\,kHz the frequency dependence is weaker which means either that the noise in our system does not %%@
have a simple $1/f^n$ dependence or that there is another noise source with a different frequency dependence that dominates at higher frequencies.  The data in fig. \ref{axialheating} were %%@
taken at the centre of the trap but we measured the same heating rates at locations $\pm270\,\mu$m along the trap $z$-axis.

We also attempted to measure the heating rate in Trap 2 but it exhibited a strong dependence on the rf drive amplitude pointing to a heating contribution from the radial modes. This, %%@
together with the large micromotion amplitude, renders a simple one-dimensional heating rate model inapplicable. We did note a large increase in heating rate without the rf filter in Trap 2; %%@
this is to be expected since an uncompensated trap is very sensitive to noise at $\Omega_{\mbox{rf}}\pm\omega_{\mbox{r}}$ \cite{Blakestad10}.  In trap 3, however, we measured no heating rate increase with variations in the out-of-plane compensation field (up to $\sim 100$\,V/m) or when running the trap without the rf filter.

\section{Future Improvements}

\subsection{Coated electrodes}
\label{sec:8a}

\begin{figure}
\begin{center}
\resizebox{0.2\textwidth}{!}{\includegraphics {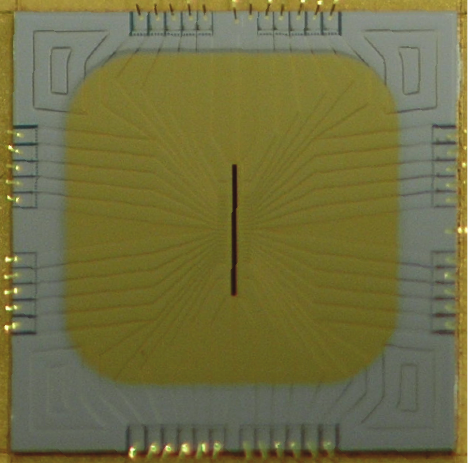}}
\end{center}
\caption{A trap where the ion trapping region has been evaporatively coated with gold.}
\label{coated}       
\end{figure}

Comparisons made by Wang and coworkers \cite{Wang11} indicate that coating the trap with another material, such as gold, which does not support a native oxide layer will reduce the charging %%@
measured in section \ref{sec:5} by more than an order of magnitude.  For this design of trap, evaporative coating is straightforward as the electrodes themselves act as a shadow mask to %%@
prevent shorting.  As this evaporation can be done as a final processing step after packaging, even materials like gold (see fig. \ref{coated}) which are not compatible with CMOS fabrication can be used.  A similar
gold coated trap has been tested and ions have been loaded into it and shuttled without problems, though detailed studies are yet to be carried out. 

\subsection{Integrated filters}
\label{sec:8b}

As demonstrated in section \ref{sec:3}, it is necessary to minimize rf pickup on control electrodes, and hence placement of capacitative filters in the closest possible proximity to the %%@
electrodes is optimal.  Additionally, filtering with a cut-off frequency larger than the DAC update rate and smaller than the rf drive and ion secular frequencies is required.  For a 1\,MHz %%@
cut-off frequency, a capacitor value of 1\,nF and a resistor value of 1\,k$\mathrm{\Omega}$ can be integrated directly into the trap chip itself as part of the chip fabrication process.  

While it is possible to monolithically integrate planar (horizontal) metal-insulator-metal plate capacitors on surface electrode traps, we find that a reliable, defect-free capacitor %%@
insulating layer using plasma deposited silicon nitride requires 30 to 50 nanometers in dielectric thickness, resulting in areal capacitance values of approximately dC/dA = 1.3 to %%@
2.1\,fF/$\mu\rm{m}^2$ and capacitor plate areas approaching 1\,mm$^2$ per electrode.  The trap chip Òreal estateÓ requirement quickly becomes prohibitively large for more %%@
than a few tens of electrodes.  Alternatives for increasing the areal capacitance values (thus decreasing the capacitor area) include using a high-k dielectric and thinning the dielectric.  %%@
The former provides quite modest improvements while the latter presents yield and reliability challenges, particularly with plasma deposited dielectrics.  

\begin{figure}
\begin{center}
\resizebox{0.35\textwidth}{!}{\includegraphics {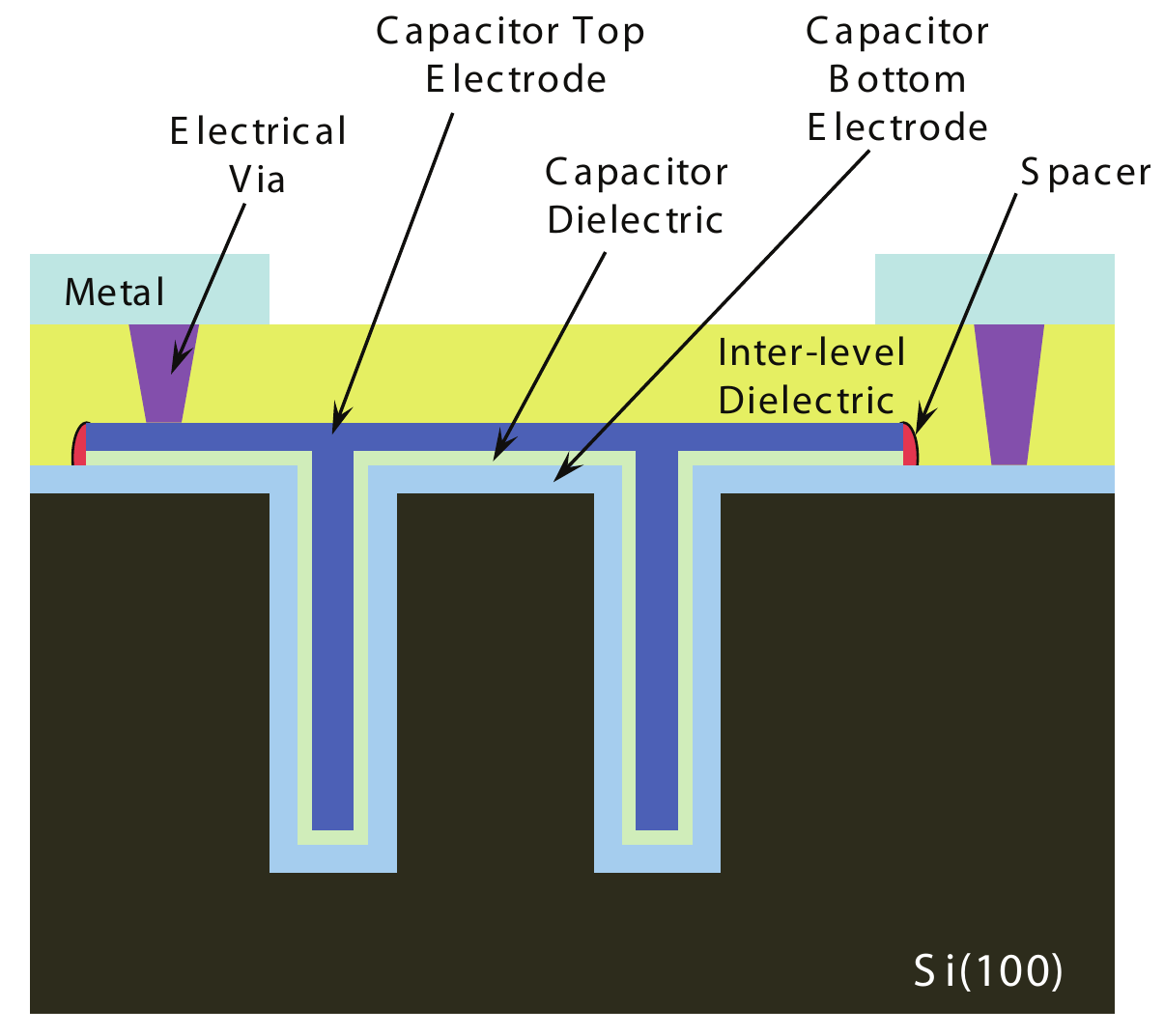}}
\end{center}
\caption{Schematic of trench capacitors built into the ion trap chip Si substrate.}
\label{trench}       
\end{figure}

A much more attractive alternative is to use low pressure chemical vapor deposited (LPCVD) dielectric films (for example stoichiometric Si$_3$N$_4$) to achieve a reliable insulator at %%@
thicknesses of 5 to 20\,nm and to create more available capacitor area by using vertical surfaces (see for example \cite{Dorp11}).  Fig. \ref{trench} shows schematically the configuration of such ``trench'' capacitors.   By %%@
building capacitors in vertical trenches etched in the Si surface of the trap chip, capacitance values on the order of 1\,nF may be achieved in horizontal (chip surface) areas of order %%@
$100\times100\,\mu\rm{m}^2$.   As an example of this, Fig. \ref{trenchgraph} below shows capacitance as a function of the chip surface area used for both trench and plate (horizontal, %%@
surface) capacitors for a 20\,nm Si$_3$N$_4$ dielectric.  Vertical trenches of width 0.5\,$\mu$m and pitch 1.0\,$\mu$m were etched 13\,$\mu$m deep in the Si substrate and phosphorus-doped %%@
LPCVD Si served as the top and bottom capacitor plates.   A factor of 30 improvement is demonstrated for the trench capacitors.  These vertical trench capacitors may be integrated with the %%@
surface electrode traps described in this work.  

\begin{figure}
\resizebox{0.48\textwidth}{!}{\includegraphics {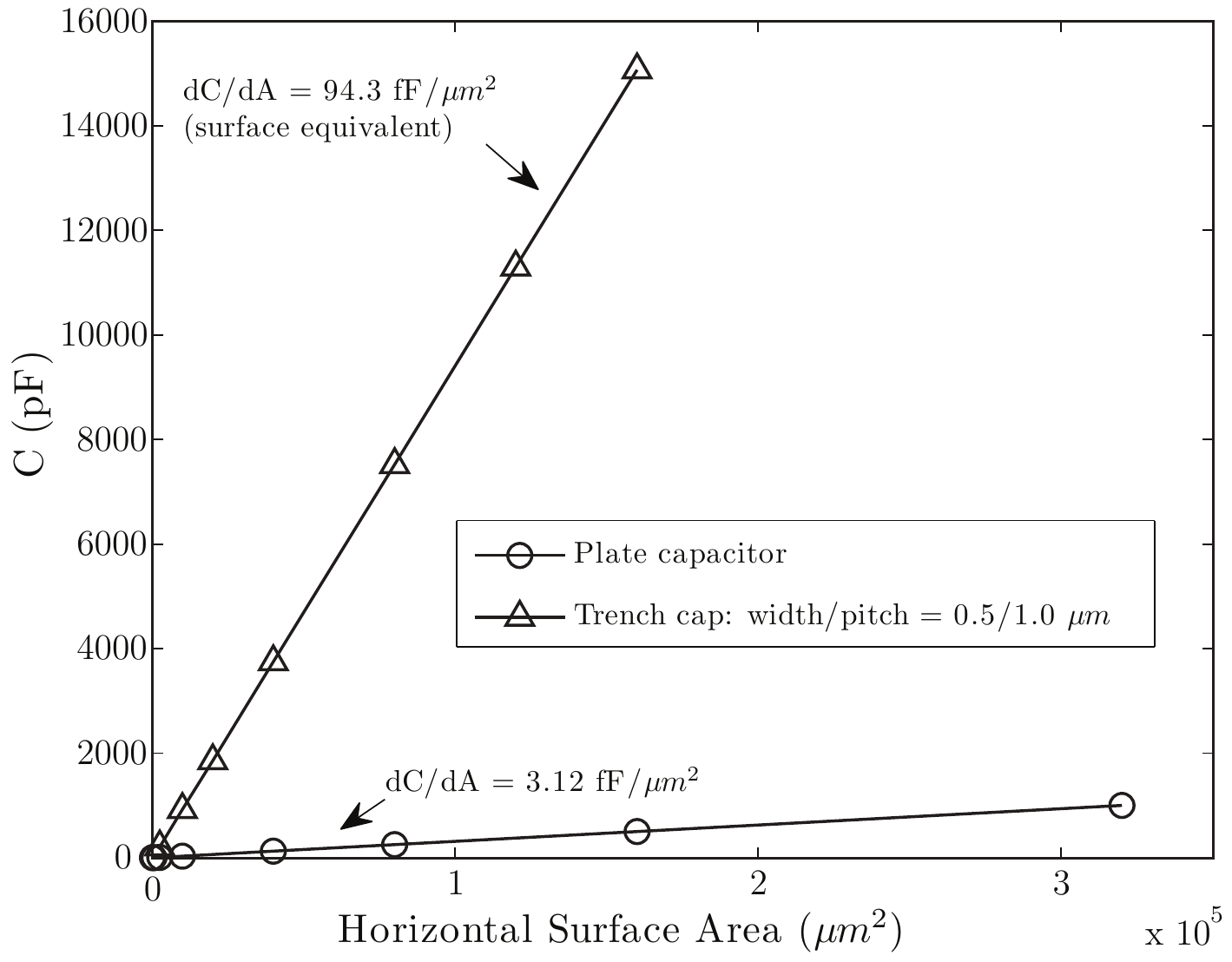}}
\caption{Capacitance values as a function of horizontal chip surface area consumed for vertical trench ($\triangle$) and surface plate ($\bigcirc$) capacitors.  The capacitor dielectric was 20\,nm LPCVD Si$_3$N$_4$.}
\label{trenchgraph}       
\end{figure}

\section{Conclusion}
\label{sec:9}

We have further characterized the performance of the microfabricated surface-electrode trap described in \cite{Stick10}, quantifying the heating rate of a single trapped ion, the charging %%@
effects on the trap by 397\,nm laser light, the ion micromotion parallel and perpendicular to the trap surface, and the ion lifetime. It was found to be essential to place capacitative %%@
filters close to the dc electrodes to prevent rf pickup and allow compensation of the ion micromotion. We described how in a future version of the design ``trench'' capacitors could be %%@
integrated on-chip to fulfil this function. We were able to load ion crystals and demonstrate sympathetic cooling of $^{43}$Ca$^+$ ions by $^{40}$Ca$^+$.

The heating rate observed in this trap is comparable to that measured in other traps of similar ion-electrode distance scale \cite{Amini08}; however, this trap has the important advantage that it is readily scalable to larger electrode arrays through the demonstrated use of junctions \cite{Moehring11}, as envisaged for an ion trap quantum information processor \cite{Kielpinsky02, Steane07}.  Furthermore, since the underlying trap device integration technique is based largely on established semiconductor integrated circuit and MEMS (micro-electro-mechanical-system) microfabrication processes, it is readily amenable to the additional integration of CMOS logic, DAC controllers and micro-optics for qubit manipulation and detection.

\begin{acknowledgement}
We are grateful to Luca Guidoni for helpful comments on the manuscript and to Derek Stacey, Graham Quelch, Jack Devlin and Martin Felle for laboratory support. This work was supported by the EPSRC Science %%@
and Innovation programme and by IARPA.  Sandia National Laboratories is a multi-program laboratory managed and operated by Sandia Corporation, a wholly owned subsidiary of Lockheed Martin %%@
Corporation, for the U.S. Department of Energy's National Nuclear Security Administration under contract DE-AC04-94AL85000.
\end{acknowledgement}

% BibTeX users please use
\bibliography{OxfordSandiaRefs}
\bibliographystyle{unsrt}

\begin{comment}

@unpublished{Linke11,
author = {N Linke et al}, 
title = {Doppler cooling with zero background using dipole transitions only},
year = {2011},
note={To be published},
%note={Also submitted for this issue},

@inbook {Amini08,
title = {Micro-Fabricated Chip Traps for Ions},
author = {Amini, J. M. and Britton, J. and Leibfried, D. and Wineland, D. J.},
author = {},
publisher = {Wiley-VCH Verlag GmbH & Co. KGaA},
isbn = {9783527633357},
url = {http://dx.doi.org/10.1002/9783527633357.ch13},
doi = {10.1002/9783527633357.ch13},
pages = {395--420},
keywords = {Ion traps, micromotion, motional heating, Paul traps, surface electrode traps, trap design},
booktitle = {Atom Chips},
year = {2011},
}

@phdthesis{Blakestad10,
author = {R Blakestad}, 
title = {Transport of Trapped-Ion Qubits within a Scalable Quantum Processor},
year = {2010},
school={University of Colorado},
}

@unpublished{Wang11,
author = {SX Wang and N Lachenmyer and Y Ge and G Low and P Herskind and I L Chuang}, 
title = {Laser-Induced Charging of Microfabricated Ion Traps},
note={Poster presented at Workshop on Ion Trap Technology 2011, Boulder CO}
}

\end{comment}

\end{document}